\documentclass{acm_proc_article-sp}

\begin{document}
\title{JASF: Jasta Security Framework \titlenote{Copyright \copyright 2011-2012 Jasta Technologies LLP.}}

\numberofauthors{2}
\author{
\alignauthor Surendranath Chowdary Chandra\\
       \affaddr{Jasta Techologies LLP}\\
       \affaddr{Bangalore,  India - 560 078}\\
       \email{surendranathchowdaryc@gmail.com}
\alignauthor Ravindranath Chowdary C\\
       \affaddr{Jasta Techologies LLP}\\
       \affaddr{Bangalore, India - 560 078}\\
       \email{ravindranathchowdaryc@gmail.com}
% 2nd. author
}

\maketitle
\begin{abstract}\textit{JASM} is a model designed to increase the security level in authentication systems. It uses IP Address of the user in the authentication process to enhance the security.\end{abstract}

 % NOT required for Proceedings

\section{Introduction}
On-line security plays a vital role to prevent users sensitive data from falling into wrong hands. JAsta Security Model(\textit{JASM}) can be used for providing higher degree of security to system/application where users authentication is required. By using this model, the problem of phishing\footnote{http://en.wikipedia.org/wiki/Phishing} can be solved to a greater extent. This model can be deployed in banking systems, mailing systems, and in high security zones. 

\section{Existing SECURITY Models}
Here we mention two most commonly used security models.

\subsection{Existing Security Model 1(\textit{\textbf{ESM1}})}
This security model typically requires a unique username and a password to login to an application. The attacker could easily get the user's credentials through phishing techniques and the user could end up being totally unaware of the event at all. Example for such systems could be email systems, where user keys in his userid and password to login to his account. 

\subsection{Existing Security Model 2(\textit{\textbf{ESM2}})}
In systems implementing \textit{ESM2}, initially user will be prompted for his username and password and these details will be sent to the application. If they are verified to be correct, then as part of the second step, the user will be prompted for verifying the One-time Verification Passcode(\textit{OTP}) sent to the user's registered mobile number with the application. On successful verification, the user will be able to login to his account. Example for such system is a 2-step verification mechanism provided by Gmail\texttrademark for (Google Inc.). \textit{ESM2} is more secure than \textit{ESM1}.

\subsubsection{Breaking of ESM2}
%\textit{ESM2} is more secure than \textit{ESM1} because it uses 2-Step 2-Channel verification. Though \textit{ESM2} is secure to some extent, it is having disadvantages. \textit{ESM2} can be easily broken, if the hacker initially get's hold of password through man-in-the-middle attack(\textit{MITM}). In this way user can initially manage to get the password of the user. In the second step, the hacker would get hold of user's mobile phone and he can get complete hold on user's account. In this way \textit{ESM2} can be completely compromised.
\textit{ESM2} is more secure than \textit{ESM1}, because it uses 2-Step 2-Channel verification. Though \textit{ESM2} is secure to some extent, it is having disadvantages. Through man-in-the-middle attack(\textit{MITM}), the attacker can easily get hold of user's password as well as his \textit{OTP}. In this way \textit{ESM2} can be completely compromised.

\section{Jasta Security Model(\textit{\textbf{JASM}})}
\textit{ESM2} depends on what user knows(password) and what user receives(\textit{OTP}) to what he has(registered mobile) to provide security. \textit{JASM} is also a 2-Step 2-Channel verification model but it attempts to increase the level of security of the authorisation systems.

%In \textit{JASM}, to enhance security, initially the user will be prompted for his username for the application. After successfully keying in of username, the user will receive \textit{OTP} on his registered mobile. Then the user will have to enter this \textit{OTP} followed by Secret Code($SC$) that he has set for the application. $SC$ is different from password. It can be user's lucky number or his favorite dish or his mobile number \textit{etc.,}. The \textit{OTP} followed by \textit{SC} is termed as \textit{OTPSC}. For example, if \textit{OTP} is $``12345"$ and $SC$ is $``ABC"$ then \textit{OTPSC} is $``12345ABC"$. After successful entry of \textit{OTPSC}, the user will be prompted for his password and upon entering his password, access will be given to his account. 

%In this system, user will enter his username for the application and he will receive a OTP on his registered mobile number. After successful entry of the passcode followed by ``$+$his registered lucky number($RLN$)"(\textit{i.e., if passcode is 656768 and his $RLN$ is 97, then he has to type in ``$656768+97$" }) he will be prompted for his password. Upon keying in his password, he will get access to his account. 

\subsection{Terminology}
\begin{itemize}
\item \textit{Secret Code(SC)}: \textit{SC} can be user's lucky number or a favorite dish or a common phrase or anything that user intends to maintain as a secret.
\item \textit{OTPSC}: \textit{OTP} followed by \textit{SC} is called \textit{OTPSC}. For example, if \textit{OTP} sent by the system is $``12345"$ and $SC$ of the user is $``ABC"$, then \textit{OTPSC} is $``12345ABC"$.
\item \textit{Account Access IP(AAIP)}: Client-side IP Address from which the request for accessing user's account was made. 
\end{itemize}

\subsection{Assumptions}
\begin{itemize}
\item There will be seamless internet connectivity
\item There will be seamless mobile connectivity
\item With each user account two mobile numbers are registered. Mobile number to which all account access related activities are to be sent by default is registered as \textit{primary mobile number}. Another mobile number is registered as \textit{secondary mobile number}, so that it can act as primary mobile number in case of loss/non-accessibility of the \textit{primary mobile number}.   
\end{itemize}

%\subsection{Flowchart}
\begin{figure*}
\centering
\includegraphics[height=7in, width=7in]{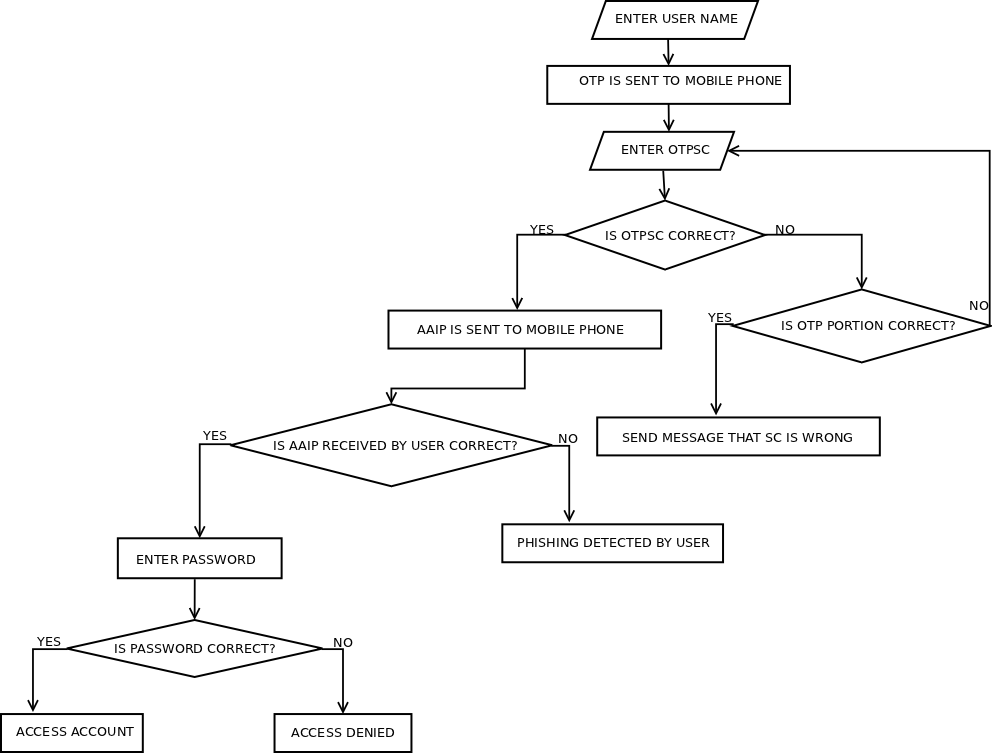}
\caption{Flowchart of \textit{JASM}}
\end{figure*}

\subsection{Description of \textbf{\textit{JASM}}}
The flowchart of \textit{JASM} is represented in \textit{Figure 1}.
\begin{enumerate}
\item User enters username to login to his account.
\item The system, sends \textit{OTP} to the user's \textit{primary mobile number} to access his account. If the user selects \textit{secondary mobile number} to receive \textit{OTP}, then the \textit{OTP} that was already sent to the \textit{primary mobile number} will be re-sent to the \textit{secondary mobile number}. 
\item User will enter \textit{OTPSC} of the corresponding account.
\item System verifies the \textit{OTPSC}.\\
Here there will be 3 possible cases.

	\begin{itemize}
		\item \textit{Case 1}: if the \textit{OTP} portion is wrong, then the system will prompt for re-entering \textit{OTPSC}.
		\item \textit{Case 2}: if the \textit{OTP} portion is correct but \textit{SC} portion is wrong, then there may be a possibility that attacker got hold of user's \textit{OTP} and he is trying to crack the \textit{SC}. System will test if such scenario is repeated for a certain threshold. If that threshold is exceeded, then it will send \textit{AAIP} stating that \textit{SC} is wrong to both the user's registered mobile numbers. Thus, it alerts the user.
		\item \textit{Case 3}: if \textit{OTPSC} is entered correctly, then \textit{AAIP} is sent to either user's \textit{primary} or \textit{secondary} mobile number based on what he has chosen. Then user verifies the correctness of \textit{AAIP} and can access his account by entering the password.
		
	\end{itemize}

\end{enumerate}

\subsubsection{Handling of OTP}
\begin{itemize}

\item User will get \textit{OTP} once he enters his username. \textit{OTP} will be valid only for a certain time-frame(\textit{e.g., 2hrs, 6hrs}). If \textit{OTP} is not used in that time-frame, user will get a new \textit{OTP} when he tries to access his account next time.
\item If someone types in other user's username, the mobile will not be flooded with \textit{OTPs}. User will get a new \textit{OTP} only after using the current \textit{OTP} or after the \textit{OTP's} time-frame expires, whichever is earlier.
\item In case of loss/non-accessibility of both the user's registered mobiles, then spl-passcodes should be available with the user as in 2-step verification mechanism provided by Gmail\texttrademark for (Google Inc.). So the user enters spl-passcode along with his $SC$. Even in this case, \textit{AAIP} will be sent to the user's registered mobile as a security measure. 
%\item Even if user's password and spl-passcodes are stolen, in \textit{JASM}, $SC$ will provide extra security and prevents the hacker from accessing the account.

\end{itemize}

\subsubsection{More of AAIP}
\begin{itemize}
\item \textit{In the absence of client-side proxy}:
The IP Address verification through \textit{AAIP} will help the user
to verify whether the \textit{OTPSC} received by the server is from his IP. \\
For \textit{e.g.}, if the user's IP is \textit{x.y.z} and he receives \textit{AAIP} as \textit{a.b.c}, then he can be sure that he is under \textit{MITM}. 
On the other hand if the \textit{AAIP} is \textit{x.y.z}, then he can be sure that he is not under \textit{MITM}. 

\item \textit{In the presence of client-side proxy}:
The IP Address verification through \textit{AAIP} will help the user
to verify whether the \textit{OTPSC} received by the server is via his proxy itself.\\
For \textit{e.g.}, if the user's proxy IP is \textit{x.y.z} and he receives \textit{AAIP} as \textit{a.b.c}, then he can be sure that he is under \textit{MITM}. On the other hand if the \textit{AAIP} is \textit{x.y.z}, then he can be sure that either himself or someone behind the same proxy have given the \textit{OTPSC} to the server. Note that in the case of a proxy, \textit{JASM} may fail if the attacker is behind the same proxy.

\end{itemize}

\section{Advantages of \textit{\textbf{JASM}} over \textit{\textbf{ESM2}}}
\begin{itemize}

\item All advantages of \textit{ESM2} are applicable to \textit{JASM}.

\item As mentioned in \textit{Section 2.2.1}, the attacker cannot get the user's password in \textit{JASM}. The reason is, the system will verify the user through \textit{OTPSC} entered by him and the user will make sure that the \textit{OTPSC} received by the server is from his IP. This is verified through \textit{AAIP} he receives. So, this mechanism acts as a \textit{2}-way handshake, before the user enters his password. So, it results in high security.
\item If the attacker gets hold of user's \textit{OTP} then also he cannot crack \textit{OTPSC}.
 
\end{itemize}

We can thus conclude that \textit{JASM} is far more secure than \textit{ESM2}.

\end{document}